\documentclass{article}

\usepackage{moreverb}

\usepackage[dvips,colorlinks,bookmarksopen,bookmarksnumbered,citecolor=red,urlcolor=red]{hyperref}

\newcommand\BibTeX{{\rmfamily B\kern-.05em \textsc{i\kern-.025em b}\kern-.08em
T\kern-.1667em\lower.7ex\hbox{E}\kern-.125emX}}

\usepackage{graphicx}
\usepackage{amssymb}
\usepackage{lineno}
\usepackage{amsbsy}
\usepackage{amsmath}
\usepackage{amsfonts}
\usepackage{graphicx}
\usepackage{epsfig}
\usepackage{multirow}
\usepackage{gensymb}
\usepackage{colortbl}
\usepackage{subcaption}

\begin{document}




\title{Improved Very-short-term Spatio-temporal Wind Forecasting using Atmospheric Classification}

\author{Jethro~Browell\footnote{Department of Electronic and Electrical Engineering, University of Strathclyde, Glasgow, UK. \texttt{jethro.browell@strath.ac.uk}},~Daniel~R.~Drew\footnote{Department of Meteorology, University of Reading, Reading, UK},~and~Kostas~Philippopoulos\footnotemark[\value{footnote}]}



\maketitle

\begin{abstract}
We present a regime-switching vector-autoregressive method for very-short-term wind speed forecasting at multiple locations with regimes based on large-scale meteorological phenomena. Statistical methods short-term wind forecasting out-perform numerical weather prediction for forecast horizons up to a few hours, and the spatio-temporal interdependency between geographically dispersed locations may be exploited to improve forecast skill. Here we show that conditioning spatio-temporal interdependency on `atmospheric modes' can further improve forecast performance. The modes are defined from the atmospheric classification of wind and pressure fields at the surface level, and the geopotential height field at the 500hPa level. The data fields are extracted from the MERRA-2 reanalysis dataset with an hourly temporal resolution over the UK, atmospheric patterns are classified using self-organising maps and then clustered to optimise forecast performance. In a case study based on 6 years of measurements from 23 weather stations in the UK, a set of three atmospheric modes are found to be optimal for forecast performance. The skill in the one- to six-hour-ahead forecasts is improved at all sites compared to persistence and competitive benchmarks. Across the 23 test sites, one-hour-ahead root mean squared error is reduced by between 0.3\% and 4.1\% compared to the best performing benchmark, and by and average of 1.6\% over all sites; the six-hour-ahead accuracy is improved by an average of 3.1\%.
\end{abstract}



\section{Introduction}
\label{sec:intro}


Wind energy is providing record a share of demand for electricity in power systems around the world and this trend is expected to continue in light of global commitments to de-carbonise society~\cite{iea2016}. Operating power systems and participating in electricity markets with high penetrations of wind energy demands continuous improvement in wind power forecasting to reduce the impact of forecast errors and uncertainty on economic cost and reliability~\cite{Skajaa2015,Foley2013}. Here we are concerned with very-short-term forecasts of the order of minutes to hours ahead which are of particular importance to participants in intra-day markets and the balancing function of power system operators~\cite{Skajaa2015,Hirth2015}.

On these time scales statistical methods based on time series analysis are generally superior to those based on post-processing numerical weather prediction due to the easy assimilation of new measurements and low computational cost of producing new forecasts~\cite{Giebel2011}. Many time series methods have been employed for wind speed and power forecasting including autoregressive (AR)~\cite{Pinson2012c} and autoregressive moving average (ARMA)~\cite{Erdem2011a} and various machine learning methods including neural networks~\cite{Li2010} and Markov chains~\cite{Carpinone2015,Yoder2013}. Hybrid methods which combine multiple prediction layers or blend forecasts from multiple methods have also been studied and shown to outperform individual methods~\cite{Feng2017}.

Information from spatially disperse measurements can be used to model spatio-temporal dependency and thereby improve forecast skill at all measurement locations~\cite{Gneiting2006,Hill2012}. Typically, measurements from multiple locations are embedded in a single vector and the temporal evolution of that vector is modelled in a vector autoregressive (VAR) framework~\cite{Lutkepohl2005}. Furthermore, the spatial dependency structure may itself depend on externalities such as season or wind direction~\cite{Hering2010,Dowell2013e}. However, the number of parameters to be estimated scales with the square of the number of spatial locations making these methods impractical for large problems. More recently, advances have been made in sparse parameter estimation in order to make large scale problems, those dealing with hundreds or potentially thousands of locations, tractable~\cite{Dowell2014c,Cavalcante2016}

In parallel to the development of spatial models, forecasting schemes based on multiple models, each designed for specific conditions, have been proposed. Regime-switching methods have been applied to forecast offshore wind power fluctuations in~\cite{Pinson2008} where the underlying regime is governed by a hidden Markov process. The number of regimes is chosen to be three by expert judgement to reflect the three distinct regions of the wind farm power curve. An adaptive extension to this approach is presented in~\cite{Pinson2012b}. A spatial regime-switching model for wind speed prediction is proposed in~\cite{Hering2010} with fixed regimes based on wind direction at a target location and selected via a cross-validation procedure. More recently, cyclone detection has been used to predict periods of potentially large forecast error in day-ahead wind power forecasting~\cite{Steiner2017}, and the EEM Wind Power Forecasting Competition was won with a regime-switching AR method with regimes identified by clustering the previous day's zonal and meridional wind speed measurements from a single location~\cite{Browell2017b}. The large-scale meteorological situation has a clear bearing on spatial dependency but to our knowledge has not been exploited in spatio-temporal time series models for very-short-term wind or wind power forecasting.

For a given region there may be a wide range of possible large-scale meteorological conditions due to variations in the strength and location of synoptic-scale weather features such as extratropical cyclones or high pressure systems~\cite{Philippopoulos2014}. Classification techniques may be used to codify these large-scale atmospheric circulation patterns in terms of a relatively small number of distinct modes~\cite{Huth2008,Philippopoulos2012} defined based on the fields of mean sea level pressure and geopotential height, for example, for each time instant of interest. Given the length-scale of the synoptic scale features, modes are typically determined on a daily temporal resolution; however, the same methods can be used to classify the patterns on any temporal scale from sub-daily to seasonal depending on the application. Here we develop a regime-switching time-series model for very-short-term forecasting with regimes defined by the atmospheric mode at each forecast issue time, with forecasts issued every hour on a rolling basis.

In this paper, we introduce a conditional regime-switching VAR model with regimes conditioned explicitly on the observed atmospheric mode at the forecast issue time. In Sections~\ref{sec:AR}--\ref{sec:paramEst} the univariate autoregressive model is introduced and extended to the proposed conditional VAR. The identification of atmospheric modes via self-organising maps is described in~\ref{sec:atclass}. A UK-based case study comprising six years of measurement data is introduced in Section~\ref{sec:casestudy} and results are presented in Section~\ref{sec:results}. Finally, conclusions are drawn and discussed in Section~\ref{sec:conc}.

\section{Forecasting Framework}
\label{sec:forecastingframework}

Consider a the wind speed time series denoted $Y=\{y_1,y_2,...,y_T\}$. We aim to forecast $y_{t+\tau}$ at time $t$ and in order to do so find some function $f_\tau(\cdot)$ which maps a vector explanatory variables $\mathbf{x}_t$ onto $y_{t+\tau}$,
\begin{equation}
	\hat{y}_{t+\tau|t} = f_\tau(\mathbf{x}_t) \quad ,
\end{equation}
while minimising some function of the prediction error, which is given by $e_{t+\tau|t}=\hat{y}_{t+\tau|t} - y_{t+\tau}$.

\subsection{Autoregression}
\label{sec:AR}
For time-series that exhibit serial correlation, such as wind speed measurements, it is reasonable for the vector of explanatory variables to consist of the recent history of $y_t$,
\begin{equation}
	\hat{y}_{t+\tau|t} = f_\tau(y_{t},y_{t-1},y_{t-2},...) \quad ,
\end{equation}
and for the function $f_\tau(\cdot)$ to take the form of a weighted sum of $p$ past values plus a constant $\nu_\tau$,
\begin{eqnarray}
	\hat{y}_{t+\tau|t} &=& \nu_\tau + \sum_{i=0}^{p-1} \alpha_{i,\tau} y_{t-i} \quad . \\
 \label{eqn:ar}
\end{eqnarray}
This is the familiar autoregressive model of order $p$, denoted AR($p$). The choice of the model order $p$ and estimation of parameters $\nu_\tau,~\alpha_{i,\tau},~i=0,...,p-1$ will be discussed in the next section. For the remainder of this section a number of extensions to the AR($p$) model specific to wind speed forecasting are introduced.

A natural extension of the autoregressive model is the inclusion of exogenous explanatory variables, sometimes denoted ARX. Since wind speed exhibits diurnal seasonality, the time of day is included as a set of dummy variables, denoted $d_h(t),~h\in\mathcal{H}$ where $\mathcal{H}$ is the set of discrete measurement times appropriate to the temporal resolution of the data, e.g. $\mathcal{H}=\{0,1,...,23\}$ in the case of hourly measurements, and
\begin{equation}
	d_h(t) =
    	\begin{cases}
        	1 & \text{if} \quad \text{Hour}(t)=h \\
            0 & \text{otherwise} \\
    	\end{cases} \quad .
\end{equation}
The autoregressive model with exogenous variables, ARX($p$), is written
\begin{eqnarray}
	\hat{y}_{t+\tau|t} &=& \sum_{i=0}^{p-1} \alpha_{i,\tau} y_{t-i} + \sum_{h\in\mathcal{H}} \beta_{h,\tau} d_h(t+\tau) \quad , 
\label{eqn:univariate+diudum}
\end{eqnarray}
where the intercept $\nu_\tau$ is superseded by $\beta_{h,\tau}$ which may be interpreted as time-dependent intercepts. Diurnal cycles may be modelled by a variety of other approaches, notably by estimating a smooth function of the time of day and de-trending the data as a form of pre-processing, as in~\cite{Hill2012}, or retaining them in an \textit{additive model}~\cite{Friedman1981} as in~\cite{Ziel2016a}. Here we proceed with dummy variables as they are flexible and easily interpretable as a time-of-day bias correction. 

Any categorical exogenous variable may be modelled in this way and we also consider ARX models where dummy variables for the current atmospheric mode $m_t \in \mathcal{S}=\{1,2,...,M\}$ are included, with associated parameters $\gamma_\tau$. In this case~\eqref{eqn:univariate+diudum} becomes
\begin{eqnarray}
	\hat{y}_{t+\tau|t} &=& \sum_{i=0}^{p-1} \alpha_{i,\tau} y_{t-i} + \sum_{h\in\mathcal{H}} \beta_{h,\tau} d_h(t+\tau) +
\sum_{s\in\mathcal{S}} \gamma_\tau \boldsymbol{1}_s(m_t)    \quad , 
\label{eqn:univariate+diudum+atdum}
\end{eqnarray}
where the indicator function $\boldsymbol{1}_s(m_t)=1$ if $m_t=s$ and $0$ otherwise. 

We conjecture that the dependence of the process $Y$ on atmospheric mode is more complex than the bias correction modelled by a dummy variable and therefore consider switching between ARX models that are mode-specific, such that~\eqref{eqn:univariate+diudum} becomes
\begin{eqnarray}
	\hat{y}_{t+\tau|t} &=& \sum_{i=0}^{p-1} \alpha_{i,\tau,m_t} y_{t-i} + \sum_{h\in\mathcal{H}} \beta_{h,\tau,m_t} d_h(t+\tau) \quad ,
\label{eqn:univariateCOND+diudum}
\end{eqnarray}
where each parameter of the ARX model depends the atmospheric mode $m_t$ at the forecast issue time.

\subsection{Vector Autoregression}

It is advantageous to consider multiple locations simultaneously in order to capture interdependency among lagged measurements for spatially dispersed sites. This is achieved by extending the AR time series models described above to vector autoregressive models whereby measurements made at time $t$ and $N$ locations and embedded in the vector $\mathbf{y}_t\in \mathbb{R}^N$ and considering the vector-valued time series $Y=\{\mathbf{y}_1,\mathbf{y}_2,...,\mathbf{y}_T\}$. The simplest VAR($p$) process is written 
\begin{equation}
	\hat{\mathbf{y}}_{t+\tau|t} = \sum_{i=0}^{p-1} \mathbf{A}_{i,\tau} \mathbf{y}_{t-i}
    \label{eqn:var}
\end{equation}
where $\mathbf{A}_{i,\tau} \in \mathbb{R}^{N\times N}$ are matrices of parameters, which is the `vectorised' form of~\eqref{eqn:ar}. Parameters on the diagonal of $\mathbf{A}_{i,\tau}$ capture autocorrelation effects and off-diagonal parameters capture cross-correlation. Exogenous variables incorporated along similar lines to give
\begin{equation}
	\hat{\mathbf{y}}_{t+\tau|t} = \sum_{i=0}^{p-1}\mathbf{A}_i \mathbf{y}_{t-i} + \sum_{h\in\mathcal{H}} \boldsymbol{\beta}_{h,\tau} d_h(t+\tau)
    \label{eqn:VAR+DD}
\end{equation}
with the effect of diurnal dummies parametrised by the vector $\boldsymbol{\beta}_{h,\tau}\in \mathbb{R}^N$, and
\begin{equation}
	\hat{\mathbf{y}}_{t+\tau|t} = \sum_{i=0}^{p-1}\mathbf{A}_i \mathbf{y}_{t-i} + \sum_{h\in\mathcal{H}} \boldsymbol{\beta}_{h,\tau} d_h(t+\tau) + \boldsymbol{\gamma}_\tau \boldsymbol{1}_s(m_t)
    \label{eqn:VAR+DD+MD}
\end{equation}
with the further addition of atmospheric mode dummies parametrised by $\boldsymbol{\gamma}_\tau \in \mathbb{R}^N$. Finally, the model parameters may themselves be dependent on atmospheric mode resulting in a conditional VAR (CVAR) model
\begin{eqnarray}
	\hat{\mathbf{y}}_{t+\tau|t} &=& \sum_{i=0}^{p-1} \mathbf{A}_{i,\tau,m_t} \mathbf{y}_{t-i} + \sum_{h\in\mathcal{H}} \boldsymbol{\beta}_{h,\tau,m_t} d_h(t+\tau) \quad ,
\label{eqn:CVAR+DD}
\end{eqnarray}
which, since $m_t$ is discrete, is equivalent to a group of $M$ VAR models, one corresponding to each atmospheric mode.

            

\subsection{Parameter Estimation}
\label{sec:paramEst}

The model parameters $\mathbf{B}_{\tau,s} = \begin{bmatrix}
    	\mathbf{A}_{1,\tau,s} & \ldots & \mathbf{A}_{p,\tau,s} &
        \boldsymbol{\beta}_{0,\tau,s} & \ldots & \boldsymbol{\beta}_{23,\tau,s} 
    \end{bmatrix}$
are estimated by minimising some function of the prediction errors on a static dataset. It is useful to define the data matrices $\mathbf{Y}_{\tau,s}$ and $\mathbf{X}_{\tau,s}$ of target and and input data, respectively, for forecast horizon $\tau$ and atmospheric mode $s$.  Input data is the vertical concatenation of explanatory variables for which a corresponding target variable exists, written
\begin{equation}
	\mathbf{X}_{\tau,s} = \begin{bmatrix}
    \vdots & & \vdots & \vdots & & \vdots \\
    	\mathbf{y}^\text{T}_i & \ldots & \mathbf{y}^\text{T}_{i-p+1} & d_0(i+\tau) & \ldots & d_{23}(i+\tau) \\
        \vdots & & \vdots & \vdots & & \vdots
    \end{bmatrix}
    \label{eqn:designMatrix}
\end{equation}
for $1 \le i < T - \tau$ subject to $m_i=s$. The corresponding matrix of target data is the vertical concatenation of wind speed vectors given by
\begin{equation}
	\mathbf{Y}_{\tau,s} = \begin{bmatrix}
    	\vdots \\
        \mathbf{y}^\text{T}_i \\
        \vdots
    \end{bmatrix}
\end{equation}
for $p+\tau \le i \le T$ subject to  $m_{i-\tau}=s$. The matrix of prediction errors for all sites and times in the dataset corresponding to mode $s$ with forecast horizon $\tau$ is given by
$\mathbf{E}_{\tau,s} = \mathbf{Y}_{\tau,s} - \mathbf{X}_{\tau,s} \mathbf{B}^\text{T}_{\tau,s}$. The parameter matrix $\mathbf{B}_{\tau,s}$ may now be estimated by minimising an appropriate function of $\mathbf{E}_{\tau,s}$.

Ordinary least squares (OLS) is used here for simplicity though different cost functions may be more appropriate for specific forecasting tasks, such as the quantile loss function for non-parametric probabilistic forecasting. The OLS parameters estimates are the solution to
\begin{equation}
	 \underset{\mathbf{B}_{\tau,s}}{\operatorname{argmin}} \|\mathbf{E}_{\tau,s} \|^2_2= 
     \underset{\mathbf{B}_{\tau,s}}{\operatorname{argmin}}
     \left\|
     	\mathbf{Y}_{\tau,s} - \mathbf{X}_{\tau,s} \mathbf{B}^\text{T}_{\tau,s}
     \right\|^2_2 \quad ,
\end{equation}
which is popular due to its simple solution by differentiation given by
\begin{equation}
	 \hat{\mathbf{B}}_{\tau,s} = 
	\left(\mathbf{X}_{\tau,s}^\text{T}\mathbf{X}_{\tau,s}\right)^{-1}
        \mathbf{X}_{\tau,s}^\text{T}\mathbf{Y}_{\tau,s} \quad .
\end{equation}
and equivalence to maximum likelihood estimation for the special case that the rows of $\mathbf{E}_{\tau,s}$ are independent and identically multivariate Normal distributed with zero mean and diagonal covariance matrix. The number of samples available for parameters estimation is an important consideration as insufficient training data will result in noisy parameter estimates~\cite{Lutkepohl2005}. As the parameters of the conditional VAR are estimated using only a subset of the available training data corresponding to a specific mode, the size of each subset may become a factor in the quality of the parameter estimates if a large number of atmospheric modes is considered, or if there is only a small amount of training data available for one or more modes.

\subsection{Atmospheric Classification}
\label{sec:atclass}

The proposed very-short-term forecasting methodology employs the large-scale atmospheric circulation as an exogenous explanatory variable and requires its classification to $M$ distinct atmospheric modes. Huth \textit{et al.}  showed there is a wide variety of methodologies available to classify circulation patterns which can be divided into three categories; subjective (manual), mixed (hybrid) and objective (automated)~\cite{Huth2008}. In this study a two-stage automated clustering approach is adopted, where the $k$-means algorithm is applied to further group the atmospheric patterns identified in the ``SOM atmospheric circulation catalogue for wind energy applications over the UK''~\cite{Philippopoulos2017}. A mixed approach is also considered whereby the second-stage grouping is made subjective judgement.

The SOM is a two-layer Artificial Neural Network (ANN) consisting of an input layer and an output two-dimensional lattice of neurons, characterized by their synaptic weights vector, $w$ and their location at the SOM lattice. Learning in SOM is achieved through the processes of competition, cooperation and adaptation. During the competition phase an input pattern is presented to the SOM and a metric distance is calculated for all neurons. The neuron with the smallest distance is the ``winner''  or \textit{Best Matching Unit} (BMU), which through a radial basis function determines the topological neighbourhood of the ``excited'' neurons at the SOM lattice during the cooperation phase. Finally, in the adaptation phase the BMU and the ``excited'' neurons' weight vectors are updated towards the input vector.

To describe the atmospheric circulation with a higher degree of generalization, a $k$-means clustering algorithm is performed as a post-processing step for further grouping the SOM patterns. The $k$-means is one of the most well known unsupervised clustering algorithms that defines the centroids through an iterative procedure and thus associates the input data to the nearest centroid. The $k$-means algorithm is analogous to the SOM learning process, with non-existing cooperation and adaptation phases. For comparison, a mix approach is tested whereby the SOM patterns are grouped by expert meteorologists through inspection of charts summarising the SOM patterns. 

\section{Case Study}
\label{sec:casestudy}

\begin{figure}
\centering
	\epsfxsize \columnwidth
	\mbox{\epsffile{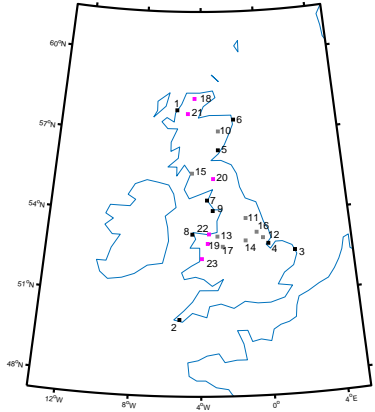}}
	\caption{Location of Measurement Sites}
	\label{fig:midasdata}
\end{figure}

The proposed method is tested on measurements of hourly mean wind speed made in the UK at 23 locations in the UK from 2002 to 2007 (as shown in Figure ~\ref{fig:midasdata}), inclusive, from the MIDAS dataset provided by the British Atmospheric Data Centre (BADC). These sites all have greater than 98\% `good' data coverage following quality control by the BADC. Years 2002--2005 are used for model order selection via 10-fold cross-validation, and 2006--2007 are used for out-of-sample testing~\cite{BADCWeb}. Forecasts are produced for 1 to 6 hours ahead for all locations meaning that in total over 2.41 million out-of-sample forecasts have been produced and evaluated. Models~\eqref{eqn:var}--\eqref{eqn:CVAR+DD} are all implemented for comparison. 

\subsection{Atmospheric Classification}

The SOM implementation used here is based on a four-step SOM clustering framework proposed in~\cite{Philippopoulos2012}, which has been also applied in a climatological context over Greece to examine the relationship of wintertime meteorological conditions with atmospheric circulation~\cite{Philippopoulos2014}. Philippopoulos~\textit{et al.} used the above framework to examine the association of atmospheric patterns with extreme wind power events in the UK~\cite{Philippopoulos2017}. The novelty of the aforementioned classification is that in addition to large-scale parameters mean sea level pressure (SLP) and geopotential height at 500hPa (Z500), the surface wind speed field (WS) is incorporated as a critical input for wind energy applications. The selected variables are extracted from Modern-Era Retrospective analysis for Research and Applications, Version 2 (MERRA-2), with an hourly temporal resolution from 1980--2014 (34 years) and for a domain centred over the UK (from 24.75$^{\circ}$W to 15.00$^{\circ}$E and 40.50$^{\circ}$N to 69.75$^{\circ}$N) with a $0.65^{\circ}\times0.5^{\circ}$ spatial resolution bilinear interpolated to a $0.75^{\circ}\times0.75^{\circ}$ grid. Upon the definition of the spatial and temporal scales (Step 1) the spatio-temporal time series are standardized and the Principal Components Analysis (PCA) is used as a pre-processing step for data reduction purposes (Step 2). The classification is performed by using the SOM algorithm on the first PC scores that explain more than one predefined percent of the initial variance, while the optimum size of the SOM feature map (the number of atmospheric patterns) is determined using qualitative criteria and the Davies-Bouldin index~\cite{Davies1979} (Step 3). The resulting catalogue of atmospheric circulation patterns may then be visualised and studied as per the user's application. The application here is to condition VAR models on atmospheric modes and that is the focus of the remainder of this paper; for more detail on the the SOM process, please see~\cite{Philippopoulos2012,Philippopoulos2014,Philippopoulos2017}.

Multiple SOM configurations for the UK were examined in~\cite{Philippopoulos2017} and the optimum size of the SOM feature map identified 21 atmospheric  organized in a 7$\times$3 map, visualised in Figure~\ref{fig:SOMgroup}. Neighbouring nodes are inter-connected and each one is associated with the composites of the selected variables. An important advantage of the approach is that relative position in the SOM map is associated with specific features, such as seasonality, location of the pressure systems and pressure gradient along with the wind field, enabling the extraction of valuable information regarding the evolution of atmospheric circulation. The results indicate that in some cases rather minor changes in large-scale atmospheric circulation may lead to a different surface wind field over the UK, a finding with important implications for wind energy applications, including forecasting. 

This study uses the hourly time-series of the 21 atmospheric patterns for the period 2002--2007. However, the SOM classification resulted in 21 atmospheric modes which left little training-data $\left\{\mathbf{X}_{\tau,s},\mathbf{Y}_{\tau,s}\right\}$ available in  for fitting each model $\mathbf{B}_{\tau,s}$, $s=1,...,21$. Furthermore it was observed that many of the 21 modes were very similar and formed natural groups. The $k$-means algorithm was used to arrange the 21 atmospheric patterns into $k$ atmospheric modes for $k=1,...,10$. In addition, two groupings were formed by expert judgement, one arranged the 21 patterns into 4 groups, and the other into 9. The grouping used in the final model was chosen via 10-fold cross validation.



\subsection{VAR Model Fitting}

Following model selection and estimation on the training dataset from 1/1/2002--31/12/2005, the performance of the proposed forecasting methodology is tested on two years of data from 1/1/2006--31/12/2007. Forecasts are produced every hour on a rolling basis from 1 to 6 hours-ahead with a dedicated model for each look-ahead time.

The autoregressive order of the VAR models~\eqref{eqn:var}--\eqref{eqn:CVAR+DD} is chosen to be $p=3$ after 10-fold cross-validation on the training dataset for values $p=1,...,5$ showed negligible difference in predictive performance, but analysis of the partial autocorrelation functions for each of the 23 series showed 3 lags to be significant in the majority of cases. While some sites showed that greater than 3 lags were significant, the results of the cross-validation exercise did not support increasing in model complexity any further.

Summary results from the cross-validation exercise are presented in Table~\ref{tab:cvresults}. Comparing the candidate models indicates that the conditional VAR based on 3 atmospheric modes has the best predictive performance across forecast horizons from 1- to 6-hours-ahead, showing greater improvement over non-conditional methods for greater forecast horizons. The conditional VAR based on the 21 patterns from the SOM (without further grouping) provides no improvement on the non-conditional VAR model.

\begin{table}
	\centering
	\caption{Root mean squared forecast error from 10-fold cross-validation on the training dataset averaged across all locations. All methods perform better then persistence across all forecast horizons, with the VAR model conditional on 3 atmospheric modes demonstrating the best predictive performance in each case.}
	\begin{tabular}{l|ccc} \hline
 		\multirow{ 2}{*}{Model}	& \multicolumn{3}{c}{RMSE (ms$^{-1}$)} \\
        		& 1-hour-ahead & 3-hour-ahead & 6-hour-ahead \\ \hline
		Persistence & 1.01 & 1.72 & 2.35 \\
        VAR      	& 0.96 & 1.55 &  2.00 \\
VAR with Diurnal Dummies& 0.94 & 1.48 & 1.87 \\
CVAR with 21 Modes 		&  0.96 & 1.49 & 1.88 \\
CVAR with 3 Modes	&  \textbf{0.93}  & \textbf{1.44} & \textbf{1.82} \\ \hline
\end{tabular}
    \label{tab:cvresults}
\end{table}

\subsection{Atmospheric Modes and Grouping}

\begin{figure}
\centering
\epsfxsize 0.7\columnwidth
	\mbox{\epsffile{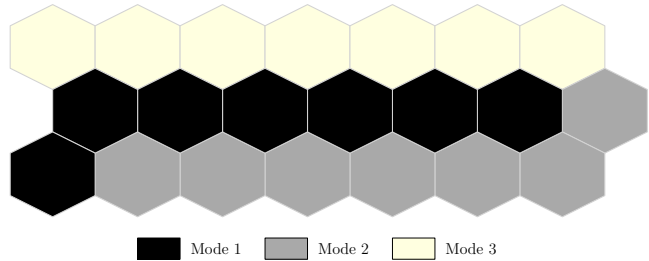}}
	\caption{Grouping of the SOM atmospheric circulation patterns and their location at the SOM map. Each hexagon represents one of the 21 atmospheric patterns identified by the SOM, shading corresponds to membership of the three clusters, or modes, found to be optimal for our forecasting application.}
	\label{fig:SOMgroup}
\end{figure}

\begin{figure}
\centering
	\epsfxsize 0.85\columnwidth
	\mbox{\epsffile{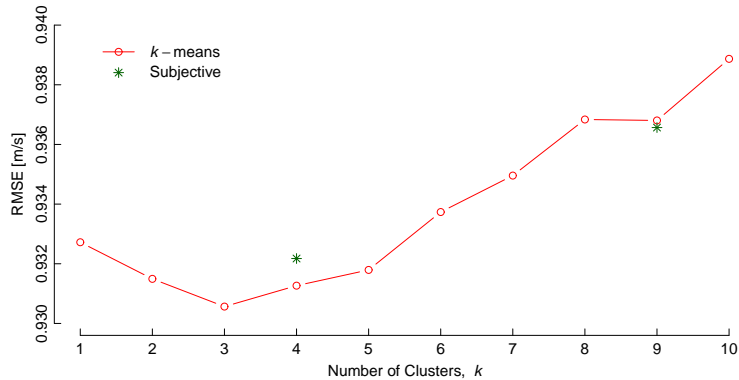}}
	\caption{Forecast performance of conditional VAR model with diurnal dummies with different groupings of atmospherics modes. Results are the product of 10-fold cross-validation on the training dataset. Different methods of forming group are considered based on clustering using all parameters (mean surface level pressure, U and V wind speed components and geopotential height at 500hPa), and subjective judgement by expert meteorologists.}
	\label{fig:groupSelection}
\end{figure}

The performance of the conditional VAR model with different numbers of atmospheric modes is plotted in~\ref{fig:groupSelection}. The special case of having 1 mode is equivalent to the standard non-conditional VAR and is outperformed by cases with two to five modes. The data-driven approach for grouping the SOM atmospheric patterns indicates that three atmospheric modes is optimal, and outperforms both groupings formed by expert judgement. The forecasting error of the 10-fold cross-validation of the conditional VAR models on the training dataset increases gradually for greater than three modes. This can be attributed to the degree of weather-related information required for improved wind forecasting skill without reducing the generalization ability of the models due to insufficient training events. Furthermore, the $k$-means grouping of the SOM atmospheric patterns is consistent with the inherent characteristic of the SOM scheme where the resulting patterns are topographically ordered in a two-dimensional map. In more detail, the first mode consists of patterns located at the second row of the SOM map, the second mode principally from the third row patterns and the third mode groups all the cases from the first row of the SOM atmospheric patterns shown in Figure~\ref{fig:SOMgroup}.

\begin{figure}	
	\centering
	\begin{subfigure}{0.3\columnwidth}
		\centering
		\epsfxsize \columnwidth
	 	\mbox{\epsffile{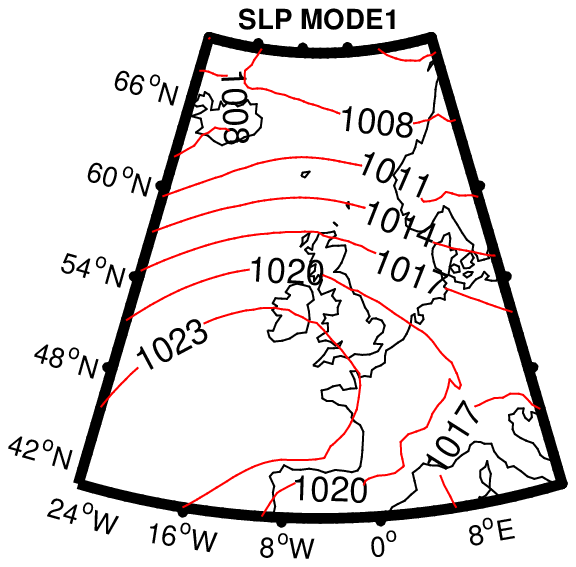}}	
	\end{subfigure}
	\begin{subfigure}{0.3\columnwidth}
		\centering
		\epsfxsize \columnwidth
	 	\mbox{\epsffile{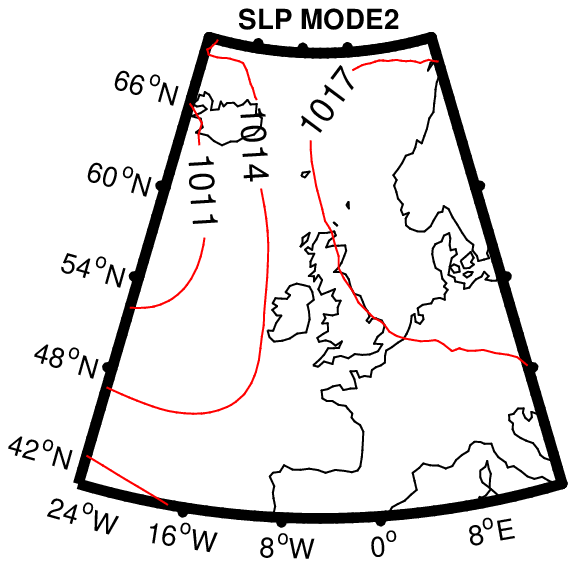}}
	\end{subfigure}
    \begin{subfigure}{0.3\columnwidth}
		\centering
		\epsfxsize \columnwidth
	 	\mbox{\epsffile{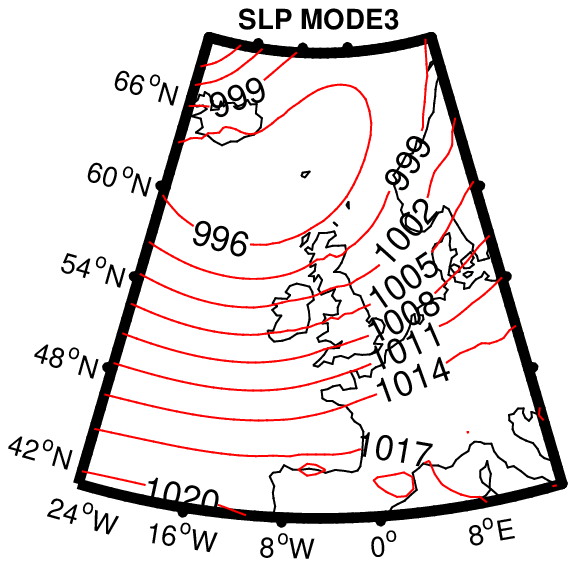}}
	\end{subfigure}

	\begin{subfigure}{0.3\columnwidth}
		\centering
		\epsfxsize \columnwidth
	 	\mbox{\epsffile{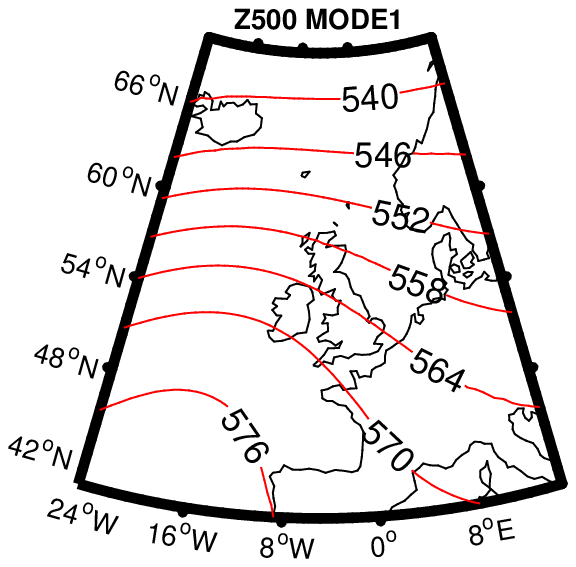}}
	\end{subfigure}
	\begin{subfigure}{0.3\columnwidth}
		\centering
		\epsfxsize \columnwidth
	 	\mbox{\epsffile{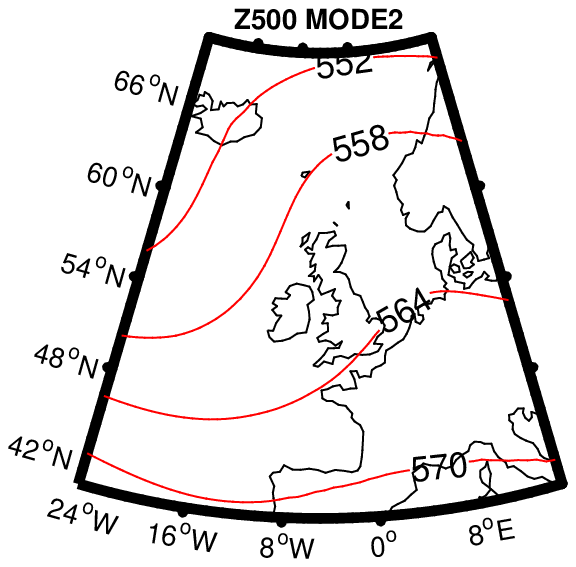}}
	\end{subfigure}
    \begin{subfigure}{0.3\columnwidth}
		\centering
		\epsfxsize \columnwidth
	 	\mbox{\epsffile{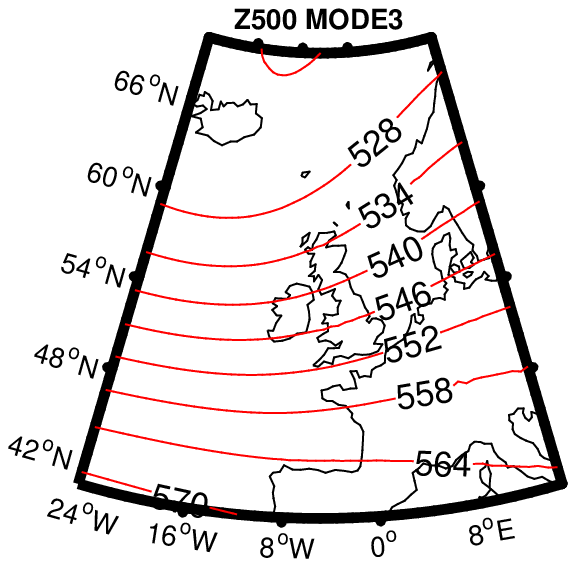}}
	\end{subfigure}

	\begin{subfigure}{0.3\columnwidth}
		\centering
		\epsfxsize \columnwidth
	 	\mbox{\hspace{0.5cm} \epsffile{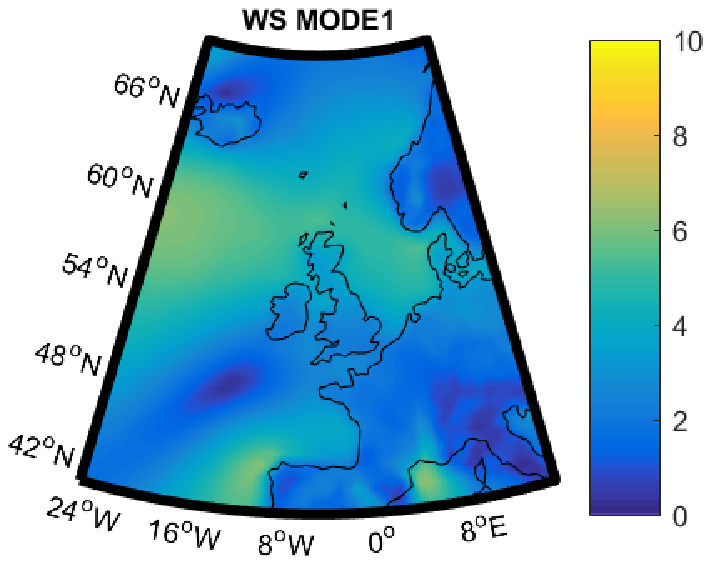}}
	\end{subfigure}
	\begin{subfigure}{0.3\columnwidth}
		\centering
		\epsfxsize \columnwidth
	 	\mbox{\hspace{0.5cm} \epsffile{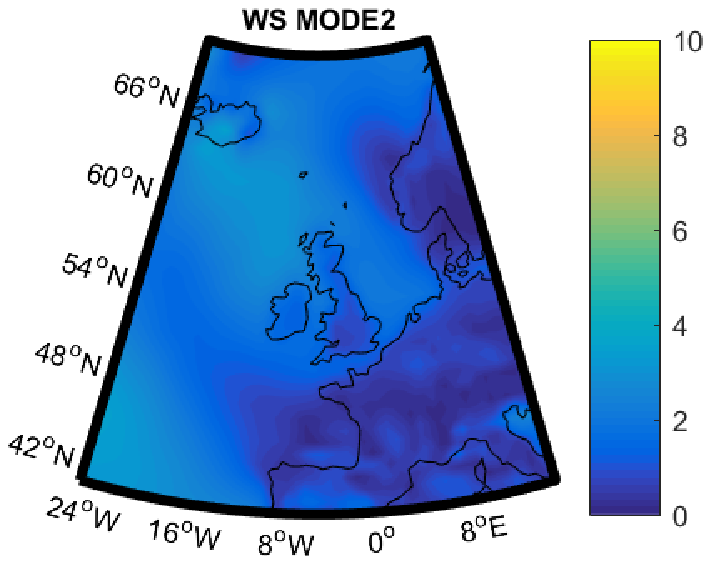}}
	\end{subfigure}
    \begin{subfigure}{0.3\columnwidth}
		\centering
		\epsfxsize \columnwidth
	 	\mbox{\hspace{0.5cm} \epsffile{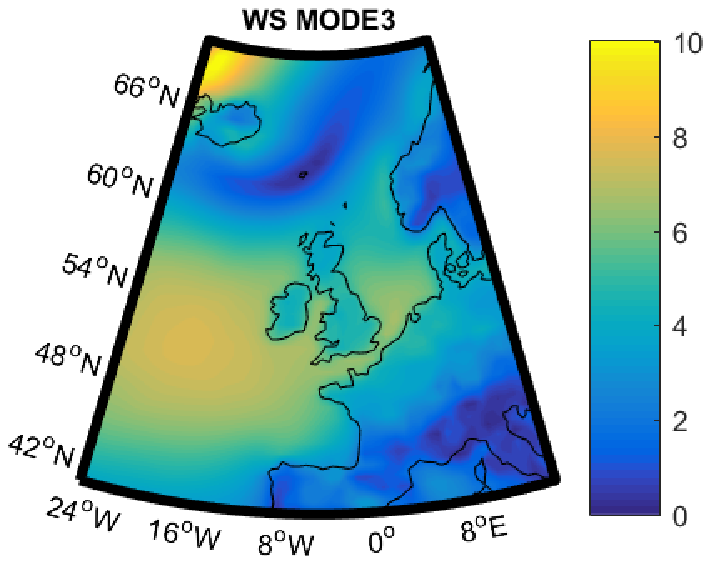}}
	\end{subfigure}

	\caption{Visualisation of surface level pressure field (SLP), geopotential height at 500hPa (Z500), and wind speed (WS) in units of ms$^{-1}$ for the three atmospheric mode centroids.}
	\label{fig:modeillustration}
\end{figure}

The modes correspond to three distinct states of atmospheric circulation and wind speed conditions over the UK. The mode centroids are illustrated in Figure~\ref{fig:modeillustration}. The first mode is associated with anticyclonic circulation and moderate wind speed conditions. The high-pressure centre is located at the south-west of UK at western Atlantic, leading to an easterly component flow with maximum intensity at Scotland and northern England. The second mode consists of low-wind speed cases and according the SLP centroid, the calm conditions observed over the UK result from the combination of the low and high pressure fields at the west and east respectively. The third mode is directly linked with the cyclonic atmospheric circulation patterns and relatively high wind speed conditions. In more detail, the centre of the low-pressure centre can either be at the west, north, east or over the British Isles and represents the frequent passage of depressions over the study area. The maximum of the westerly component surface flow is observed over central and southern areas of the UK.

\begin{figure}
\centering
\epsfxsize \columnwidth
	\mbox{\epsffile{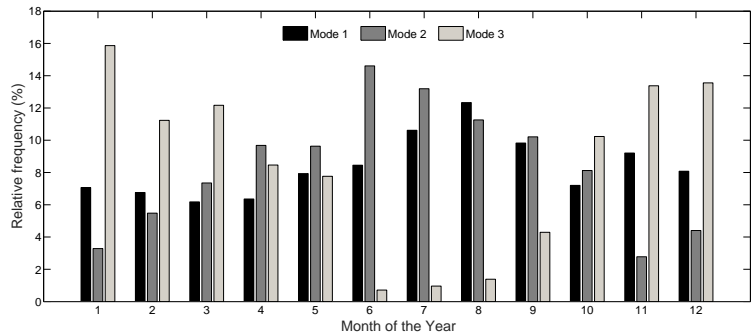}}
	\caption{Relative frequency of occurrence for each atmospheric mode by month.}
	\label{fig:groupMonthFreq}
\end{figure}

For the study period 2002--2007, the atmospheric conditions are evenly distributed across the three modes (32.4\% of the time in Mode 1, 36.0\% of the time in Mode 2 and 31.6\% of the time in Mode 3). Both Modes 1 and 2 occur throughout the year, however there is a higher frequency of events during the summer period, as shown in Figure~\ref{fig:groupMonthFreq}. Approximately 40\% of Mode 1 events occur between May and August where the high-pressure is related to the extension of the Azores anticyclone. Similarly, approximately 50\% of mode 2 events are between May and August. However, in comparison to Mode 1, Mode 2 occurs less frequently during the winter months, less than 13\% of events occur in between December and February. In contrast, Mode 3 is more common during the winter and transitional seasons, less than 5\% of events occur between June and August. This is due to the frequent passage of extratropical cyclones associated with the North Atlantic storm track.

\begin{figure}
\centering
 	\epsfxsize 0.85\columnwidth
	\mbox{\epsffile{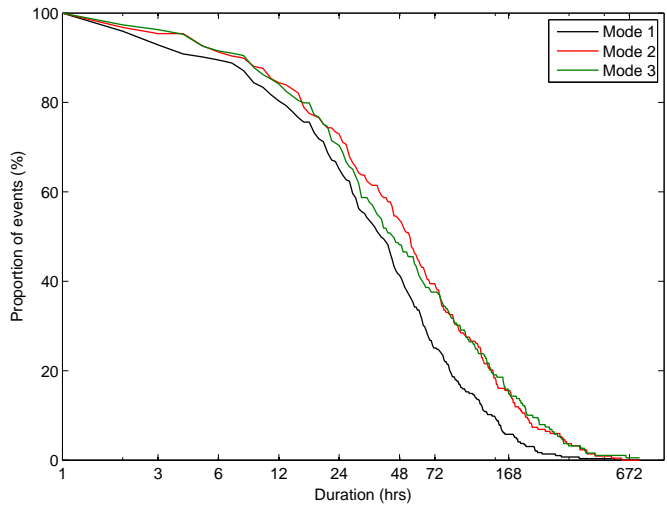}}
	\caption{Frequency distribution of the duration for which  each mode event persists during the period 2002--2007.}
	\label{fig:mode_cdf}
\end{figure}

The atmospheric modes are defined by the synoptic scale conditions, which have a length scale of the order of 1000km and therefore tend to persist for a time period of days. Figure ~\ref{fig:mode_cdf} shows the frequency distribution of the duration for which each mode event persists during the period 2002--2007. The distributions of Modes 2 and 3 are very similar; the mean duration is 86 and 87 hours respectively (which equates to 3.6 days), in comparison the duration of Mode 1 is generally shorter (mean duration of 57 hours or 2.4 days). This difference is largely due to the extremely long duration events which occur more frequently for Modes 2 and 3. The median duration for Mode 1 (38 hours) is actually very similar to Mode 3 (42 hours), however Mode 3 is much more likely to persist for very long periods (over 10 days). This is generally associated with the passage of consecutive extratropical cyclones which are sufficiently close to prevent a change in the atmospheric mode. For all three modes it is very rare for the duration of an event to be shorter than 6 hours (10.5\% for Mode 1, 8.7\% for Mode 2 and 8.5\% for Mode 3).

The underlying physical process governing the transition between atmospheric modes is clearly non-Markovian as the passage of synoptic-scale weather features is smooth and occurs on a temporal scale much greater than the hourly sample rate under consideration. The probability of transitioning from one mode to another will depend on more than the just present state (the definition of a Markov process) and will depend, for instance, on the amount of time spent in the present state. This supports our efforts to observe the atmospheric mode rather than taking a Markov-switching type approach based on an assumption that the transition between regimes is governed by a hidden-Markov model. 

\section{Results}
\label{sec:results}

\begin{figure}
\centering
	\epsfxsize 0.85\columnwidth
	\mbox{\epsffile{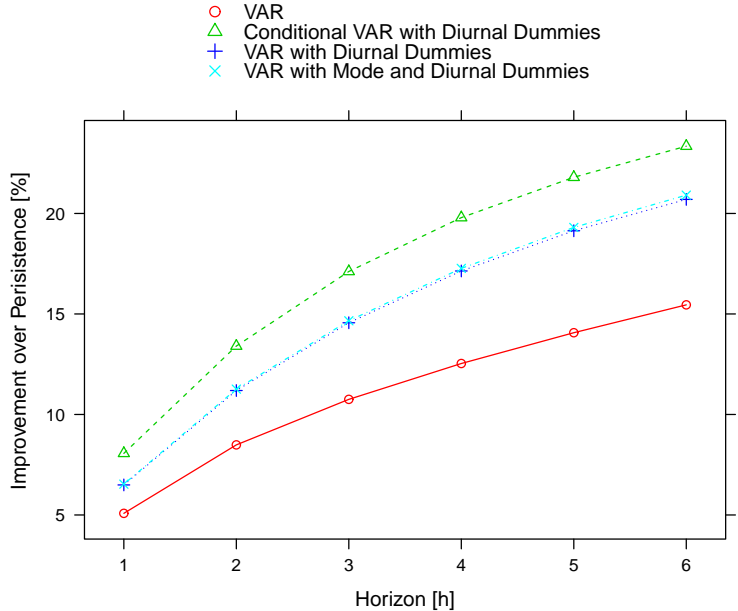}}
	\caption{Improvement over persistence for averaged across all sites on test dataset.}
	\label{fig:imp}
\end{figure}

Conditioning the forecast on atmospheric modes increases the accuracy of the wind speed predictions across sites and all lead times. For the 1-hour ahead forecast the  RMSE of the predicted wind speed averaged across the 23 sites is reduced by 1.6\% relative to VAR with Diurnal Dummies and 7.8\% relative to persistence, as illustrated in in Figure~\ref{fig:imp}. The improvement in the forecast skill due to the information provided by the atmospheric modes increases with the lead time. For example, for the 6-hour ahead forecast, CVAR improves the RMSE by 3.1\% relative to VAR with diurnal dummies, and 23.9\% relative to persistence. Figure~\ref{fig:imp} also shows that adding mode dummies to the VAR model with diurnal dummies has very little impact on the skill of the model. This indicates that the additional skill provided by the CVAR model is due to the better representation of the spatial structure of winds between the sites provided by the atmospheric modes, it is not simply a bias correction.

\begin{figure}
\centering
	\epsfxsize \columnwidth
	\mbox{\epsffile{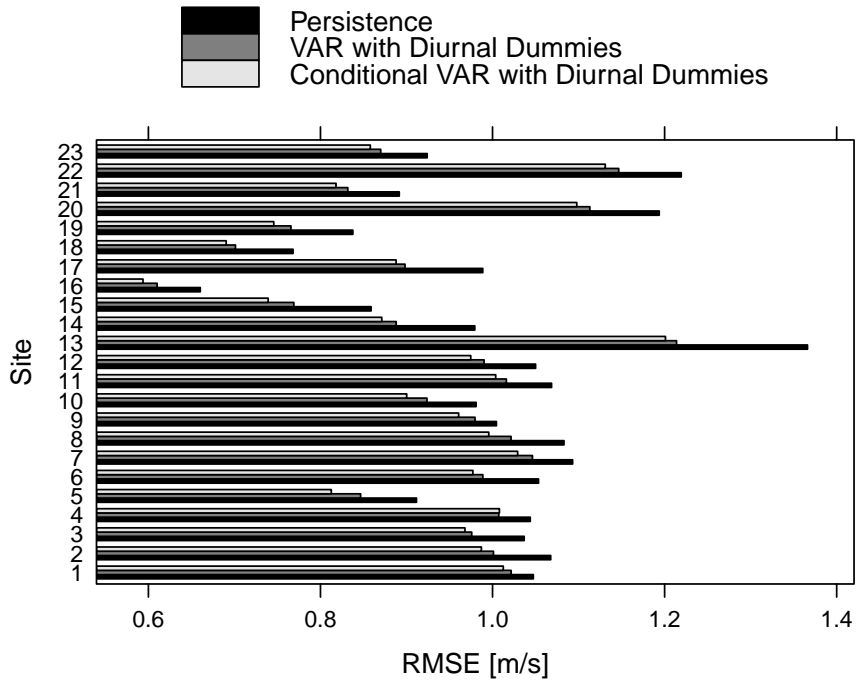}}
	\caption{Variation in error across sites for 1-hour-ahead forecast comparing Persistence, VAR and Conditional VAR.}
	\label{fig:site_rmse}
\end{figure}

The CVAR model produces an improvement in the forecast at all 23 sites (see Figure ~\ref{fig:site_rmse}); however the magnitude of the reduction in RMSE varies from site to site. For example, for the 1-hour-ahead forecasts the reduction in RMSE varies from only 0.35\% for Site 4 to 4.1\% for Site 5. This result is also true for all of the atmospheric modes, i.e. for all sites there is a reduction in the RMSE of the wind speed using the CVAR model for each of the atmospheric modes. In general, the CVAR model provides the greatest improvement in the forecast when the atmosphere is determined to be in Mode 3, cyclonic conditions. Averaged across the 23 sites, there is a reduction in the RMSE of 2.4\% during Mode 3 events, in comparison to a reduction of 1.1\% and 1.4\% for Modes 1 and 2, respectively. Furthermore, the reduction in RMSE is greatest for Mode 3 for 16 of the 23 sites. Further analysis of the sites has not revealed a clear relationship between the added value due to the atmospheric modes and the geographical location, terrain type and the elevation of the sites. 

\begin{figure}
\centering
	\epsfxsize \columnwidth	\mbox{\epsffile{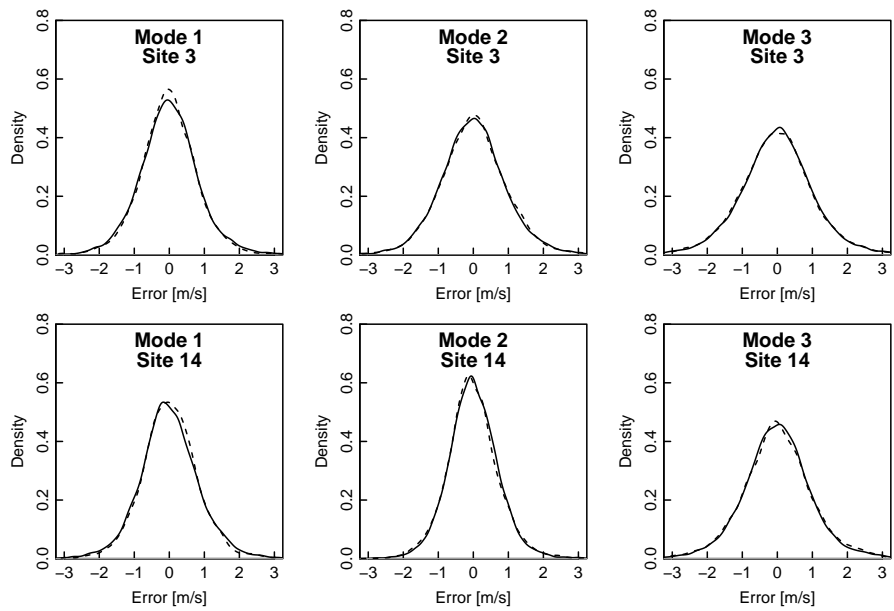}}
	\caption{This distribution of forecast errors separated by atmospheric mode for training data (solid line) and test data (dashed line). Illustrated sites, top: Gorleston, bottom: Nottingham.}
	\label{fig:residComp}
\end{figure}

The distribution of the forecast errors is approximately Gaussian for all 3 atmospheric modes. Figure~\ref{fig:residComp} shows the distributions for two locations, however similar results were shown for all of the other sites. The spread of the distributions varies for each mode and location. For all sites, the largest errors occur for Mode 3 which is not surprising given it is associated with cyclonic conditions and relatively high wind speeds. For the majority of the sites the distribution of the errors for Modes 1 and 2 are very similar. However, for the a number of the sites in Scotland and northern England the errors are larger for Mode 1, when these sites tend to experience higher wind speeds, as shown in Figure~\ref{fig:modeillustration}.


\section{Conclusion}
\label{sec:conc}

This paper presents a framework for incorporating the information provided by the large-scale meteorological conditions into  a vector-autoregressive method for very-short-term wind forecasting using an atmospheric mode classification. The approach has been applied to a case study based on 6 years of measurements from 23 locations across the UK. As a result of the information  provided by the atmospheric modes on the spatial and temporal structure of the wind field, the forecast skill is improved at all sites and lead times compared to competitive benchmarks. For the 1-hour ahead forecast, the RMSE of the predicted wind speed averaged across the 23 sites is reduced by 1.6\% relative to VAR with Diurnal Dummies and 7.8\% relative to persistence. 

An improvement in forecast skill was shown at all 23 sites for all atmospheric modes. However, the model generally provided the greatest improvement in the forecast during cyclonic conditions (Mode 3). Averaged across the 23 sites, there was a reduction in the RMSE of 2.4\% during Mode 3 events, in comparison to a reduction of 1.1\% and 1.4\% for Modes 1 and 2, respectively. For each mode, the distribution of the forecast errors was approximately Gaussian at all 23 sites. The spread of the distributions however varies for each mode and location. Despite, the increased forecast skill, the wind speed errors were typically largest for mode 3 (cyclonic conditions).

While forecast performance is consistently improved when atmospheric conditions were grouped into 3 modes, grouping conditions into 6 or more modes was detrimental to forecast accuracy, compared to non-conditional methods. Given the size of the dataset, it is unlikely that this is due to having insufficient data for parameter estimation alone. The unsupervised learning approach used for atmospheric classification presented here is not fully optimised for forecast performance, rather groupings are formed based on generic distance metrics. Semi-supervised learning methods may enable atmospheric classification to be performed in such a way that groupings are formed to explicitly improve forecast performance. 

The framework presented in this study can be applied to any geographical location or combination of sites, however there are several key areas of consideration. Firstly, the atmospheric classification has been performed using reanalysis data. To run operationally, the method could be adapted to determine the atmospheric mode from the analysis of forecasts provided by a Numerical Weather Prediction model. Secondly, the application of this method to a large number of sites should consider sparse VAR approximation. The sparsity structure of such models could provide further insight into the nature or spatio-temporal structures under different atmospheric conditions. Finally, at present the model has only been applied to wind speed forecasting, therefore further work is required to quantify the benefits for wind power forecasting.

\section*{Acknowledgements}
We gratefully acknowledge the British Atmospheric Data Centre for their provision of the MIDAS dataset of meteorological measurements, and the North American Space Administration for the provision of the MERRA-2 dataset. Jethro Browell is supported by the University of Strathclyde's EPSRC Doctoral Prize, grant number EP/M508159/1. 

\textbf{Data Statement:} The reanalysis data from the used in this study is free to download from https://gmao.gsfc.nasa.gov/reanalysis/MERRA-2/, UK meteorological observations are freely available to bona fide research programmes from~\cite{BADCWeb}. Data and code produced in the course of this work are available to download from the University of Strathclyde KnowledgeBase~\cite{BrowellData2017a}.

\bibliographystyle{plain}
\bibliography{atmos_class.bib}







\end{document}